\documentclass[12pt]{article}

\usepackage{hyperref}
\usepackage{amssymb}

\begin{document}

\renewcommand{\abstractname}{\hfill}

\newpage
\pagenumbering{arabic}

{\Large \bf Particles with Negative Energies in Nonrelativistic

\vspace{4pt}
and Relativistic Cases
}

\vspace{9pt}
{\bf
Andrey A. Grib${}^{1,2,*}$ and Yuri V. Pavlov${}^{3,4}$
}
\begin{abstract}
${}^{1}$ Theoretical Physics and Astronomy Department, The Herzen  University,
48~Moika, St.\,Petersburg, 191186, Russia

${}^{2}$ A. Friedmann Laboratory for Theoretical Physics, St.\,Petersburg, Russia

${}^{3}$ Institute of Problems in Mechanical Engineering of
Russian Academy of Sciences,
61 Bolshoy, V.O., St.\,Petersburg, 199178, Russia; yuri.pavlov@mail.ru

${}^{4}$ N.I.\,Lobachevsky Institute of Mathematics and Mechanics,
Kazan Federal University, Kazan, Russia

${}^{*}$ Correspondence: andrei\_grib@mail.ru
\end{abstract}
\vspace{-27pt}
    \begin{abstract}
\noindent
{\bf Abstract:}

    States of particles with negative energies are considered for the
nonrelativistic and relativistic cases.
    In nonrelativistic case it is shown that the decays close to the attracting
center can lead to the situation similar to the Penrose effect for rotating
black hole when the energy of one of the fragments is larger than the energy of
the initial body.
    This is known as the Oberth effect in the theory of the rocket movement.
    The realizations of the Penrose effect in the non-relativistic case in
collisions near the attracting body and in the evaporation of stars from
star clusters are indicated.
    In relativistic case similar to the well known Penrose process in the
ergosphere of the rotating black hole it is shown that the same situation as
in ergosphere of the black hole occurs in rotating coordinate system in
Minkowski space-time out of the static limit due to existence of negative energies.
   In relativistic cases differently from the nonrelativistic ones the mass of
the fragment can be larger than the mass of the decaying body.
    Negative energies for particles are possible in relativistic case in
cosmology of the expanding space when the coordinate system is used with
nondiagonal term in metrical tensor of the space-time.
    Friedmann metrics for three cases: open, close and quasieuclidian, are analyzed.
    The De Sitter space-time is shortly discussed.

\vspace{7pt}
\end{abstract}
\vspace{-7pt}
    \noindent
\hrulefill

\section{\normalsize Negative Energies and the Penrose Effect
in Nonrelativistic Case}
\label{secNRS}

\hspace{\parindent}
    It is well known that in the non-relativistic case, energy is determined
with accuracy to the additive constant.
    If the energy of a particle resting on infinity is put equal to zero
then the sign of the energy defines movement of the particle
in Kepler problem being either limited (in the case of negative sign)
or nonlimited, when the energy is positive or zero~\cite{LL_I}.
    If the body arriving in the region with nonzero gravitational field of
some other object decays there in two parts so that the velocity of one
part is smaller than the second cosmic velocity in the point of decay then
its energy will be negative.
    So the energy of the second part becomes larger than the energy of the
initial body.
    This is some realization of the Penrose
effect~\cite{Penrose69,PenroseFloyd71} on getting the energy from the
rotating black hole due to the decay of some body in the ergosphere.

    Let us give some evaluations for the case when the initial body
with mass~$m$ and the velocity $v_0$ on infinity decays on the distance~$r$
from the attracting center of mass~$M$ on two fragments so that the fragment
with mass $m_1$ is flying relative to the first part of mass $m-m_1$ in the
opposite direction to the initial one with the velocity $u$.
    The velocity of the initial body at the distance~$r$ from the attracting mass
$M$ is
    \begin{equation}    \label{Un1}
v_r = \sqrt{v_0^2 + \frac{2 G M}{r} },
\end{equation}
    where $G$ is the gravitational constant.
    Let us find the velocities of the fragments after the decay using
the conservation of the momentum.
    So one gets for the velocity of fragment with mass $m-m_1$
    \begin{equation}    \label{Unv2}
v_2 = v_r + \frac{m_1}{m} u .
\end{equation}
    The projection of the velocity of the fragment with mass $m_1$ on
the direction of the initial movement in the point of decay is
    \begin{equation}    \label{Unv1}
v_1 = v_r - \left( 1 -  \frac{m_1}{m} \right) u.
\end{equation}
    It is easy to prove the identity
    \begin{equation}    \label{UnTE}
E_2 + E_1 = E_0 + E_f,
\end{equation}
    where
    \begin{equation}    \label{UnTE2}
E_2 = (m - m_1) \left( \frac{v_2^2}{2} - \frac{G M}{r} \right)
\end{equation}
    is the energy of the fragment with mass $m-m_1$ in the gravitational
field of the attracting body.
    \begin{equation}    \label{UnTE1}
E_1 = m_1 \left( \frac{v_1^2}{2} - \frac{G M}{r} \right)
\end{equation}
    is the energy of the fragment with mass  $m_1$,
    \begin{equation}    \label{UnTE0}
E_0 = \frac{m v_0^2}{2}
\end{equation}
    is the kinetic energy of the initial body far from from the attracting body.
    Note that in the nonrelativistic case one has in the equation of energy
conservation for the considered process~(\ref{UnTE}) the term
    \begin{equation}    \label{UnTEf}
E_f = \frac{m_1 u^2}{2}.
\end{equation}
    This is the energy necessary for the flight of the fragment with the
relative velocity $u$.
    It is evident interpretation that $m_1$ is the mass of the fuel flying from the nozzle
of the rocket and $E_f$ is the energy of the fuel.

    As it is seen from~(\ref{UnTE}) in case $E_1- E_f <0$ one has the analogy
with the Penrose process.
    The full mechanical energy of the fragment with mass $m-m_1$ (the rocket
without fuel) is larger than the initial mechanical energy of the initial object $E_0$.
    From~(\ref{Unv2})--(\ref{UnTEf}) one can find the conditions of
the realization of such process
    \begin{equation}    \label{UnTEp}
E_2 - E_0 = E_f - E_1 = m_1 \left[ v_2 u \left( 1- \frac{m_1}{m} \right) -
\frac{v_0^2}{2} \right].
\end{equation}
    This expression is positive for small $r$.
    Note that formally the profit in energy is unlimited, however it is evident
that $r$ must be larger than the Schwarzschild radius $r_g = 2 G M /c^2$
where $c$ is the light velocity.
    As one can see from~(\ref{UnTEp}) to get the energy profit comparable with
the relativistic rest mass of the fuel $m_1 c^2$ one needs relativistic values of the
fuel expiration velocity and use of gravitational field close to
the gravitational radius.
    Surely in this case one must do calculations using relativistic theory
(change the Newton potential on the Schwarzschild metric and taking into
account that relative velocities are close to the velocity of light).

    Note that for $m_1 \ll m$ and $v_0^2 \ll v_2 u$ formula~(\ref{UnTEp}) shows
that the profit in the energy of the rocket in the engine start with
the big velocity in the region of movement is explained by the fact that
the engine traction with  the fuel expiration relative velocity $u$
is constant and the work $ A $ done in this case is proportional to
the velocity of the rocket movement:
    \begin{eqnarray}    \label{UnA}
A \approx \frac{u m_1}{\Delta t} v_2  \Delta t = v_2 u m_1.
\end{eqnarray}

    It was H.\,Oberth~\cite{Oberth1929} in 1929 who was the first to propose
this way to increase the efficiency of the rocket engine by start of the engine
in the periaster of the trajectory.
    It is used in astronautics in gravitational maneuvers using the Moon
and inter planet flights.

    Another example of realization of the Penrose effect in nonrelativistic case
is the following.
    Let on two meeting circular orbits around the body with attracting
mass $M$ two particles 1, 2 with masses $m_1 \gg m_2$ are rotating on the
distance~$r$.
    Then in case of absolute elastic collision the particle with
mass $m_2$  will move from the first particle with relative
velocity $ 2 \sqrt{GM/r}$.
    Its velocity in the system of rest will be $ 3 \sqrt{GM/r}$
and so the energy will be $7 GM m_2/(2r)$ increasing in $4 GM m_2/(r)$.
    It is evident that the energy of the first particle decreases in the same
value.
    Note that the profit in energy of the order close to $m_2 c^2$ is
possible only near the gravitational radius of the attracting body.

    If three or more particles are interacting one also can have process
similar to the Penrose process.
    For example such decays occur for three interacting stars
or evaporation of stars clusters~\cite{Syunyaev}.

\vspace{4mm}
\section{\normalsize  Negative Energies in Rotating Coordinates}
\label{secRNE}
\hspace{\parindent}
    In relativity theory negative energies are usually absent.

    In nonrelativistic limit in the Kepler problem of the massive body
with mass $m$ moving around gravitating mass $M$ on the distance $r$
the full energy of this body taking into account the rest mass energy
can be less than zero if this distance is very small
    \begin{equation}    \label{nr1}
E = m c^2 + \frac{m v^2}{2} - G \frac{mM}{r} <0 \ \ \Rightarrow \ \
r < \frac{G M}{c^2} = \frac{r_g}{2}.
\end{equation}
    Here $r_g$ is the gravitational radius.

    However physics in the ergosphere of the rotating black hole shows
that as it was discovered by R.\,Penrose~\cite{Penrose69} relativistic
particles can have negative energies due to dependence of the energy
on the angular velocity of the body rotating around the black hole.
    In our papers~\cite{GribPavlov2016d,GribPavlov2019} it was shown that
similar effect occurs in rotating coordinates in Minkowski space
out of the static limit defined by us in analogy with the ergosphere.

    The interval in Minkowski space in rotating cylindrical coordinates is
    \begin{equation}    \label{v3}
d s^2 = (c^2 - \Omega^2 r^2)\, dt^2 + 2 \Omega r^2 d \varphi\, d t -
d r^{2} - r^{2} d \varphi^{2} - d z^{2},
\end{equation}
    where $\Omega$ is the angular rotation velocity.

    Let us consider this formula for the case of our solar system
when the Earth in at rest (nonrotating around its axis) --- as ancient
Greeks thought.
    The use of such rotating coordinate system is important because
after all our observatories and observers are at rest in this case.
    If ${\Omega_\oplus \approx 7.29 \cdot 10^{-5}}$\,s$^{-1}$
is the angular rotational velocity then from~(\ref{v3}) one has
$g_{00}=0$ for $ r_s = c/\Omega_\oplus = 4.11 \cdot 10^9$\,km.
    It is some distance between the orbits of Uranus and Neptune.
    For distance $r> r_s$ the coefficient of metric $g_{00} <0$
but this does not mean that $ds^2 <0$ because of the presence of
nondiagonal term.
    For $ds^2 >0$ physical motion with the velocity $v<c$ is still
possible and no breaking of causality occurs.
    We call the distance $r_s$ the static limit in analogy of
the corresponding surface in the Boyer-Lindquist coordinates
of rotating black hole~\cite{BoyerLindquist67}.
    For $r> r_s$ no body can remain at rest in the rotating
coordinate system.

    In cylindrical coordinates of Minkowski space 4-vector of energy-momentum
    \begin{equation}    \label{v6d}
p'_{\, i} = \left( \frac{E'}{c}, \ - p^{\prime\,r},
\ - L'_z, \ - p^{\prime z} \right),
\end{equation}
    where
    \begin{equation}    \label{v6kk}
L'_z = r^2 p^{\prime \varphi} = m r^2 \frac{d \varphi' }{d \tau}
= \frac{E'}{c^2} r^2 \frac{d \varphi' }{d t}
\end{equation}
    is the projection of the angular momentum on the~$z$ axis.
    Transforming~(\ref{v6d}) to the rotating coordinates one obtains
    \begin{equation}    \label{v10kk}
p_{i} = \left( \frac{E' + \Omega L'_z}{c},
\ - p^{\prime\,r}, \ - L'_z, \ - p^{\prime z} \right).
\end{equation}
    Therefore, the energy in these coordinates is
    \begin{equation}    \label{v10v}
E = E' + \Omega L'_z.
\end{equation}
    So the energy $E$ can be negative depending on the sign of $L'_z = L_z$.

    As it is known the energy can be written with the help of
the Killing vector~$\zeta^i$ as
    \begin{equation} \label{NEnergy2}
E^{(\zeta)} = \int_\Sigma T_{ik}\, \zeta^i \, d \sigma^k ,
\end{equation}
    where~\{$\Sigma$\} is some set of spacelike hypersurfaces
orthogonal to~$\zeta^i$
and $T_{ik}$ is the energy-momentum tensor of some matter.
     Note that negative values of the energy $E^{(\zeta)}$ can be obtained for the
positive energy density $T_{ik}$  if the Killing vector becomes spacelike
as it is the case for the ergosphere of the rotating black hole and in
our case in rotating coordinates in region  out of the static limit.
    This means that the conditions of the Penrose-Hawking theorem on singularities
in cosmology are not broken in spite of existence of particles with negative energies
considered in this paper.
    Another remark concerns negative energy of the galaxy on the surface of
the expanding sphere with homogeneous density of matter inside it resembling
the closed Universe in Newton's approximation.
    In exact relativistic case it is not the energy but the energy density of matter
is present in Einstein's equations.
    Energy is not conserved in expanding Universe.
    However as one can see in next part of our paper this non conserved energy
of the particle can have in some situations negative sign.

    For a pointlike particle with the mass~$m$ located at point $x_p$
one obtains (see~\cite{GribPavlov2016d})
    \begin{equation} \label{NEnergy5}
T^{ik}(x) = \frac{m c^2 }{\sqrt{|g|}} \int \! ds \, \frac{d x^i}{d s}\,
\frac{d x^k}{d s}\, \delta^4 ( x - x_{p})
\end{equation}
    and
    \begin{equation} \label{NEnergyEgen}
E^{(\zeta)} = m c^2\, \frac{dx^i}{ds}\,  g_{ik} \zeta^k = c (p, \zeta).
\end{equation}
    For the rotating coordinates~(\ref{v3}) and $\zeta = (1,0,0,0)$
we obtain the energy $E^{(\zeta)}$ is equal to~(\ref{v10v}).

    One can easily obtain the condition for a negative energy beyond the static
limit~\cite{GribPavlov2019}
    \begin{equation}    \label{Novu}
L_z < - \sqrt{\frac{p_z^2 c^2 + m^2 c^4}{ \Omega^2 - (c/r)^2 }}, \ \ \ \
\frac{v}{c} > \frac{c}{r \Omega}, \ \ \ \
r> c/\Omega .
\end{equation}

    Some experiment to observe the consequences of the Penrose process
for our solar system was proposed in~\cite{GribPavlov2019}.
    The idea of the experiment is to show the possibility of such processes
which can be interpreted differently in two different coordinate systems ---
the inertial one and the noninertial rotating coordinates.
    The particle beyond the static limit decays on two fragments
one with the positive energy, the other with the negative one.
    The observers on the Earth can not see the particle with negative energy
because it exits only out of the static limit.
    But he can see the fragment with positive energy which in rotating
coordinates is larger than the energy of the initial decaying particle.
    From the inertial point of view the energies of all particles are positive
but the energy of particle falling on the Earth is larger than the energy
of the initial particle if one takes into account the rotation of the Earth
around its axis.
    All quantitative estimates one can find in~\cite{GribPavlov2019}.

    Further we shall discuss the third possibility of existence of
particle states with negative energy --- that of the expanding Universe.

\vspace{4mm}
\section{\normalsize Negative Energies and Static Limit in Expanding Universe}
\label{secNEEU}

\hspace{\parindent}
    Expanding Universe in the standard model is described in
the Friedmann-Robertson-Walker form in synchronous frame~\cite{LL_II} as
    \begin{equation}
d s^2 = c^2 d t^2 - a^2(t)
\left( \frac{d r^2}{1 - k r^2} + r^2 d \Omega^2 \right),
\label{f1}
\end{equation}
    where $d \Omega^2 = d \theta^2 + \sin^2 \theta \, d \varphi^2 $,
$k = 1 $ for closed cosmological model,
$k=-1$ for open model and $k=0$ for quasi-Euclidean flat model.

        The metric~(\ref{f1}) can be written also in other coordinates
    \begin{equation}
d s^2 = c^2 d t^2 - a^2(t)
\left( d \chi^2 + f^2(\chi) d \Omega^2 \right),
\label{f1n}
\end{equation}
    where
$$
f(\chi) = \left\{
\begin{array}{ll}
\sin{\chi} , & \ \ \ k=1, \\
\chi , & \ \ \ k=0, \\
\sinh{\chi} , &\ \ \ k=-1
\end{array}
\right.
$$
    under replacing $r= f(\chi)$.
    In closed model $\chi$ is changing from 0 to $\pi$,
in cases $k=0,-1$ one has $\chi \in [0, + \infty)$.
    The radial distance between points $\chi=0$ and $\chi$
in metric~(\ref{f1n}) is $D= a(t) \chi$
and it's the same in the metric~(\ref{f1}).
    If $t$ is fixed  then the maximal value of $D$ is $D_{\rm max} =\pi a(t)$.
    In open and flat models $D$ is non limited.

    Take the new coordinates $t, D, \theta, \varphi$
(see also~\cite{Ellis93,Grib95}).
    Then
    \begin{equation}
d D = \frac{\dot{a}}{a} D\, d t + a \, d \chi , \ \ \ \
d \chi = \frac{1}{a } \left( d D - \frac{\dot{a}}{a}\, D d t \right)
\label{f3}
\end{equation}
    and the interval~(\ref{f1n}) becomes
    \begin{equation}
d s^2 = \left( 1 - \left( \frac{\dot{a} }{a } \frac{D }{c } \right)^2 \right) c^2 d t^2
+ 2 \frac{\dot{a}}{a} D\, d D d t - d D^2 - a^2 f^2(D/a)\, d \Omega^2 .
\label{f4}
\end{equation}
    The interval~(\ref{f4}) is the special case of the more general interval
    \begin{equation}
d s^2 = g_{00} (dx^0)^2 + 2 g_{01} dx^0 dx^1 + g_{11} (dx^1)^2 + g_{\Omega \Omega} d \Omega^2,
\label{f5}
\end{equation}
    where $g_{11} < 0 $, $ g_{\Omega \Omega} < 0 $ and $g_{00} g_{11} -g_{01}^2 < 0 $.

    As it was shown in our paper~\cite{GribPavlov2020} for the energy
of the particle one has
    \begin{equation}
E= p_0 c = mc^2 g_{0k} \frac{d x^k}{ d s} =
mc^2 \frac{d x^0}{ d s} \left( g_{00} + g_{01} \frac{d x^1}{ d x^0} \right).
\label{f7}
\end{equation}
    This energy due to limitation from causality $ds^2 >0$
    \begin{eqnarray}
\frac{g_{01} - \sqrt{ g_{01}^2 - g_{11} g_{00} - g_{11} g_{\Omega \Omega}
\left( \frac{d \Omega}{ d x^0 } \right)^2 } }{-g_{11}}
\le \frac{d x^1}{d x^0}  \le
\frac{g_{01} + \sqrt{ g_{01}^2 - g_{11} g_{00} - g_{11} g_{\Omega \Omega}
\left( \frac{d \Omega}{ d x^0 } \right)^2 } }{-g_{11}}
\label{f6}
\end{eqnarray}
    is limited by inequality
    \begin{eqnarray}
        \nonumber
\frac{ m c}{-g_{11}} \frac{ d x^0 }{d \tau }
\left( g_{01}^2 - g_{11} g_{00} - |g_{01}| \sqrt{ g_{01}^2 - g_{11} g_{00} -
g_{11} g_{\Omega \Omega} \left( \frac{d \Omega}{ d x^0 } \right)^2 }\right)
\le E \le \\
\frac{ m c}{-g_{11}} \frac{ d x^0 }{d \tau }
\left( g_{01}^2 - g_{11} g_{00} + |g_{01}| \sqrt{ g_{01}^2 - g_{11} g_{00} -
g_{11} g_{\Omega \Omega} \left( \frac{d \Omega}{ d x^0 } \right)^2 }\right).
\label{f8}
\end{eqnarray}
      As one can see the states with zero and negative energy are possible
 in the region where $g_{00} <0$ (out of the static limit).
     In the chosen coordinate system in this region no physical body
 can be at rest.
    In this region movement will be observed either from the observer
if  $g_{01} > 0$  corresponding to the expanding Universe~(\ref{f4})
with  $\dot{a}>0$ or to the observer for  $g_{01} < 0$ in case
of contracting Universe~(\ref{f4}) with $\dot{a}<0$.

    The ``static'' limit for the metric~(\ref{f4}) is
    \begin{equation}
D_s = \frac{c }{|h(t)|},
\label{fps}
\end{equation}
    where $ h(t) = \dot{a}/a $ is the Hubble parameter.
    This corresponds to the radius of the so-called light
sphere~\cite{Ellis93}.
    In open and quasi-Euclidean models due to non limited $D$ the region
always exists with $D> D_s$, where particles can't be at rest in used
coordinates and states with negative energies are possible.
    In closed model $D\le \pi a(t)$ and the condition for the existence of
such region is
    \begin{equation}
|\dot{a}(t)| > c / \pi .
\label{n2}
\end{equation}

    To understand the meaning of the limit~(\ref{n2}) consider
closed Universe with dust matter.
        Then (see~\cite{LL_II})
    \begin{equation}
a=a_0 ( 1 - \cos \eta), \ \ \
t= \frac{a_0}{c} ( \eta - \sin \eta), \ \ \ \eta \in (0, 2 \pi), \ \ \
a \in (0, 2 a_0),
\label{n3}
\end{equation}
    $D_s = a | \tan (\eta /2 )| $ and~(\ref{n2}) becomes
    \begin{equation}
\left| \tan \frac{\eta}{2} \right| < \pi.
\label{n4}
\end{equation}
    So that in region $ \eta \in (0, 2 \pi) $ one has
    \begin{equation}
0 < \eta < 2 \tan^{-1} \pi, \ \ \ \
2 (\pi - \tan^{-1} \pi) < \eta < 2 \pi.
\label{n5}
\end{equation}
    The scale factor in region~(\ref{n5}) is in the limits
    \begin{equation}
0 < a < \frac{2 \pi^2a_0 }{1 + \pi^2}
\label{n6}
\end{equation}
    and
    \begin{equation}
D_s = \frac{D_{\rm max}}{\pi} \left| \tan \frac{\eta}{2} \right| \le D_{\rm max},
\label{n6d}
\end{equation}
    i.e. the region of existence of negative energies in closed dust
Unverse is changing from the almost entire Friedman universe except of
the observer vicinity at $\eta \to 0, 2 \pi$ to empty set at $\eta $
outside the intervals~(\ref{n5}).

    Note that for closed dust Universe for $\eta \ll 1$ one has
    \begin{equation}
a \approx \left( \frac{9 c^2 a_0}{2} \right)^{1/3} t^{2/3}.
\label{n7}
\end{equation}
    As it seen from~(\ref{n5}) for this case one has the region
out of the static limit with states with negative energies.

    Static limit in the considered model at $\eta \lesssim 0.74 \pi$
 lies in the domain of the particle's horizon
    \begin{equation}
l_p = a(t) \int_0^{t} \frac{d t'}{a(t')} = a \eta.
\label{npH}
\end{equation}
    This means that the region with negative particle energies intersects with
the region of causally connected phenomena for the observer at the origin of
coordinate system.

    The energy of the freely moving particle is
    \begin{equation}
E= E' \left( 1 - \frac{\dot{a}^2}{a^2 } \frac{D^2}{c^2} +
\frac{\dot{a}}{a} \frac{D}{c^2} \frac{d D}{d t} \right),
\label{f9}
\end{equation}
    where $E' = m c^2 dt / d \tau $.
   So in these coordinates particles with negative energies
are moving in coordinate $D > D_s$ so that the velocity is slower
than some definite value
    \begin{equation}
\frac{d D}{ d t} < \frac{ c^2 a}{D \dot{a} } \left( \frac{D^2}{D_s^2} - 1 \right).
\label{f10}
\end{equation}
    Let us rewrite the inequality~(\ref{f10}) in terms of coordinates
$t, \chi, \theta, \varphi$.
    Then we obtain in the case $\dot{a} > 0$
    \begin{equation}
v = a \frac{d \chi}{d t} < - c \frac{D_s}{D}, \ \ \ D > D_s.
\label{vfon}
\end{equation}
    This has a meaning similar to that obtained by us for
the case of rotating coordinate frame $t, \chi, \theta, \varphi$:
particles with negative energies close to the static limit  ($D \to D_s$)
must move with velocities close to the light velocity in direction
of the observer in expanding Universe.

    The necessary condition for the possibility of observing processes
involving particles with negative energy is $D_s < L_H$,
where
    \begin{equation}
L_H = a(t) c \int \limits_{t}^{t_{\rm max}} \frac{d t'}{a(t')}
\label{aKHn}
\end{equation}
is cosmological event horizon,
$t_{\rm max}$ is the life time of the universe.
    In the case of the de Sitter Universe with the scale factor
$a = a_0 \exp H t$, where $a_0$ and $H$ are constants, $ t_{\rm max} =\infty$,
the Hubble constant is constant and the static limit $D_s = c/H$
is constant and it is equal to the cosmological event horizon.
    So processes (Penrose processes) with particles with negative energy
are not seen by the observer in this case.
    This is analogous to situation of nonrotating black holes where particles
with negative energy exist only inside the event
horizon~\cite{GribPavlov2010NE}.
    Note that in the limit $t \to \infty$, the standard cosmological
$\Lambda \rm CDM$ model tends to the de Sitter stage.

    However for models with $k=0$ and scale factors
$a = a_0 t^\alpha$ ($ 0 < \alpha <1 $) the cosmological horizon $L_H = \infty$
and some visible consequences of existence of particles with negative energies
can be observed.
    In the closed Universe with dust matter~(\ref{n3})
$t_{\rm max} = 2 \pi a_0/c $ and we have
    \begin{equation}
L_H = a(\eta) (2 \pi - \eta).
\label{aKHnn}
\end{equation}
    The condition $D_s < L_H$ is reduced to
    \begin{equation}
\left| \tan \frac{\eta}{2} \right| < 2 \pi - \eta.
\label{nz}
\end{equation}
    This inequality in particular is true over the entire interval of
the existence of the static limit in the era of expansion of
the closed universe $0 < \eta < 2 \tan^{-1} \pi$
and also at the end of the compression era.
    Thus  some processes with the particles with negative energies
can be observed in the models of closed universe also.
    However, unlike the de Sitter universe, the energy of
the particle is not conserved in these cases.
This makes the manifestation of the Penrose effect less obvious.

\vspace{7pt}
\noindent
{\bf Author Contributions:}
The authors equally contributed to this research work.

\vspace{7pt}
\noindent
{\bf Funding:}
This research is supported by the Russian Foundation for Basic research (Grant No. 18-02-00461 a).
The work of Yu.V.P. was supported by the Russian Government Program of Competitive Growth
of Kazan Federal University.

\vspace{7pt}
\noindent
{\bf Conflicts of Interest:}
The authors declare no conflict of interest.


\end{document}